\documentclass[aps,prc,reprint,showpacs,groupedaddress]{revtex4-1}
\usepackage{amssymb}
\usepackage{graphicx}
\usepackage{color}

\begin{document}

\newcommand{\ApJ}{Astrophys. J.}
\newcommand{\MNRAS}{MNRAS}
\newcommand{\AandA}{Astron. Astrophys.}
\newcommand{\be}{\begin{equation}}
\newcommand{\ee}{\end{equation}}
\newcommand{\Chandra}{{\it Chandra}\ }
\newcommand{\densnuc}{\rho_{\mathrm{nuc}}}
\newcommand{\nb}{n_{\mathrm{b}}}
\newcommand{\nnuc}{n_{\mathrm{nuc}}}
\newcommand{\Tcnt}{T_{\mathrm{cnt}}}
\newcommand{\Tcp}{T_{\mathrm{cp}}}
\newcommand{\Teff}{T_{\mathrm{eff}}}
\newcommand{\Tc}{T_{\mathrm{c}}}
\newcommand{\kf}{k_{\mathrm{Fx}}}
\newcommand{\kfn}{k_{\mathrm{F}n}}
\newcommand{\kfp}{k_{\mathrm{F}p}}
\newcommand{\Msun}{M_{\mathrm{Sun}}}
\newcommand{\Mdu}{M_{\mathrm{dU}}}
\newcommand{\chisqr}{\chi_\nu^2}
\newcommand{\zg}{z_{\mathrm{g}}}
\newcommand{\Ts}{T_{\mathrm{s}}}
\newcommand{\Tsinfty}{\Ts^{\infty}}

%-------------------------------------------

\title{Tests of the nuclear equation of state and superfluid and superconducting gaps using the Cassiopeia A neutron star}
\author{Wynn C.~G. Ho}
\email{email: wynnho@slac.stanford.edu}
\affiliation{Mathematical Sciences and STAG Research Centre, University of Southampton, Southampton, SO17 1BJ, United Kingdom}
\author{Khaled G. Elshamouty}
\author{Craig O. Heinke}
\affiliation{Department of Physics, University of Alberta, CCIS 4-181, Edmonton, AB, T6G 2E1, Canada}
\author{Alexander Y. Potekhin}
\affiliation{Ioffe Institute, Politekhnicheskaya 26, 194021 Saint Petersburg, Russia}
\affiliation{Central Astronomical Observatory at Pulkovo, Pulkovskoe Shosse 65, 196140 Saint Petersburg, Russia}
% \date{\today}
\date{Received 5 August 2014; revised manuscript received 4 December 2014; published 14 January 2015}

%-------------------------------------------
\begin{abstract}
The observed rapid cooling of the Cassiopeia A neutron star can be interpreted
as being caused by neutron and proton transitions from normal to
superfluid and superconducting states in the stellar core.
Here we present two new \Chandra ACIS-S Graded observations of this neutron
star and measurements of the neutron star mass $M$ and radius $R$ found from
consistent fitting of both the X-ray spectra and cooling behavior.
This comparison is only possible for individual nuclear equations of state.
We test phenomenological superfluid and superconducting gap models which
mimic many of the known theoretical models against the cooling behavior.
Our best-fit solution to the Cassiopeia~A data is one in which the
($M$,$R$) = ($1.44\,\Msun$,$12.6\mbox{ km}$) neutron star is built with the
BSk21 equation of state, strong proton superconductor and moderate neutron
triplet superfluid gap models, and a pure iron envelope or a thin carbon
layer on top of an iron envelope,
although there are still large observational and theoretical uncertainties.
\end{abstract}

\pacs{97.60.Jd, 26.60.-c, 67.10.-j, 95.85.Nv}

\maketitle

\section{Introduction \label{sec:intro}}

The study of neutron stars (NSs) provides a unique probe of the nuclear
equation of state (EOS), which prescribes a relationship between pressure
and density and determines the behavior of matter near and above nuclear
densities ($\nnuc\approx 0.16\mbox{ fm$^{-3}$}$ or
$\densnuc\approx 2.8\times 10^{14}\mbox{ g cm$^{-3}$}$).
Current theories indicate that the core of NSs
(at $n\gtrsim 0.1\mbox{ fm$^{-3}$}$)
may contain a neutron superfluid and proton superconductor and exotic
particles, such as hyperons and deconfined quarks, may exist in the inner
core (at $\rho\gg\densnuc$)
(see, e.g., \cite{haenseletal07,lattimer12}, for review).
The EOS also determines the total mass $M$ and radius $R$ of a NS,
and therefore measurements of $M$ and $R$ can be used to infer
the EOS \cite{haenseletal07,lattimer12,lattimersteiner14}.
One example where $M$ and $R$ are measured is for the NS in the Cassiopeia~A
(Cas~A) supernova remnant.
By fitting \Chandra X-ray spectra of this source with theoretical models,
the best-fit mass and radius are found to be $M=1.62\,\Msun$ and
$R=10.2\mbox{ km}$ \cite{elshamoutyetal13};
the flux energy spectra depends on mass and radius through the brightness
(function of $R^2$), gravitational redshift (function of $M/R$), and surface
gravity (function of $M/R^2$ and redshift), with the last having a relatively
weak effect on spectra.
The allowed ranges of values for the Cas A NS are not particularly
constraining, i.e., $M\approx 1.3-2\,\Msun$ and $R\approx 8-15\mbox{ km}$
\cite{hoheinke09,yakovlevetal11}.

A complementary method for uncovering the EOS, as well as other fundamental
physics properties, is by investigating the cooling behavior of NSs.
NSs begin their lives very hot (with temperatures $T>10^{11}\mbox{ K}$)
but cool rapidly through the emission of neutrinos.
The processes that govern neutrino emission depend on physics at the
supra-nuclear densities of the NS core.
Importantly, unlike energy spectra which depend only on the bulk properties
of the NS (such as $M$ and $R$)
and its surface properties, the cooling behavior depends critically on
details of the EOS, e.g., neutron and proton number densities
(see \cite{tsuruta98,yakovlevpethick04,pageetal06}, for review).
For the case of the Cas~A NS, measurement of rapid cooling
\cite{heinkeho10,shterninetal11,elshamoutyetal13}
provides the first constraints on the critical temperatures for the onset
of superfluidity of core neutrons $\Tcnt$ (in the triplet state) and protons
$\Tcp$ (in the singlet state),
i.e., $\Tcnt\approx(5-9)\times 10^8$~K and $\Tcp\sim (2-3)\times 10^9$~K
\cite{pageetal11,shterninetal11}.

However these critical temperature constraints are obtained assuming either
the (X-ray spectra) best-fit mass \cite{shterninetal11} or
a fit to the temperature decline by varying $M$ but neglecting whether this
value of $M$ (and implied $R$) leads to a good fit of the spectra
\cite{pageetal11}.
Here we fit the temperature evolution of the Cas~A NS with particular EOSs
and superfluid and superconducting energy gaps at the same time as evaluating
how well the mass and radius predicted by the EOS fits the X-ray spectra.
In other words, for each EOS, we determine the quality of the spectral fit
along the $M$-$R$ sequence predicted by that EOS; we then use that EOS to
calculate the cooling behavior and test whether this theoretical behavior
matches the observed behavior.
To do this fully consistently,
a complete NS model requires a self-consistent calculation of the EOS and
superfluid and superconducting gap energies.
However, this has not been done up to the present time.  Therefore we assume
that the EOS and gap models are decoupled,
as in \cite{pageetal04,pageetal09}.
We also assume standard (i.e., minimal) cooling
\cite{gusakovetal04,pageetal04}, since cooling by fast neutrino emission
processes, such as direct Urca, produces temperatures that are far too low
at the current age of the Cas~A NS ($\sim 330\mbox{ yr}$; \cite{fesenetal06}).
With these assumptions, we perform for the first time consistent fitting
of both the Cas~A NS spectra and temperature evolution for the NS mass and
radius.
We find that the mass and radius can be determined very accurately for a
given EOS and gap energies.
However there are sufficient observational and theoretical uncertainties
that we cannot claim to rule out specific EOS and gap energy models.
One of the main purposes of this work is to motivate nuclear physicists
to not only calculate the EOS, but also superfluid and
superconducting gap energies, and to provide them in a useful way to the
astrophysicists.

In Sec.~\ref{sec:obs}, we discuss our new observations of the Cas~A NS.
In Sec.~\ref{sec:model}, we briefly describe our NS
model, including the EOS and superfluid and superconducting gaps.
In Sec.~\ref{sec:results}, we present our results.
Finally, we summarize and discuss our conclusions in Sec.~\ref{sec:discuss}.

\section{Cas~A temperature data, including new \Chandra observations
\label{sec:obs}}

The two new data points are from 49-ks and 50-ks ACIS-S Graded observations
taken on 2013 May 20 (ObsID 14480) and 2014 May 12 (ObsID 14481), respectively.
We use the Chandra Interactive Analysis of Observations (CIAO) 4.5 software
and Chandra Calibration Database (CALDB) 4.5.5.1 to analyze all the ACIS-S
Graded observations.
For each observation, we calculate ancillary response functions,
including corrections for the fraction of the point-spread function
enclosed in an extraction region.
We fit all the spectra simultaneously to measure NS surface temperatures
using the non-magnetic partially ionized carbon atmosphere models of
\cite{hoheinke09}, adopting the same fitting parameters as in
\cite{shterninetal11,elshamoutyetal13}, and holding the NS mass and radius,
distance, and hydrogen column density fixed between observations.
Further details are described in \cite{elshamoutyetal13}
(see also \cite{posseltetal13}).
The results are shown in Table~\ref{tab:tempgraded}.
Note that in the present work, we consider the rapid cooling rate derived
from only these ACIS-S Graded data; future work will consider the lower
cooling rates found by \cite{elshamoutyetal13,posseltetal13}.

%-------------------------------------------
\begin{table}
\caption{\Chandra ACIS-S Graded mode temperatures. \label{tab:tempgraded}}
\begin{ruledtabular}
\begin{tabular}{ccc}
ObsID & Year & $\Teff$\footnotemark[1] \\
\hline
114 & 2000.08 & $2.145_{-0.008}^{+0.009}$ \\
1952 & 2002.10 & $2.142_{-0.008}^{+0.009}$ \\
5196 & 2004.11 & $2.118_{-0.007}^{+0.011}$ \\
(9117,9773)\footnotemark[2] & 2007.93 & $2.095_{-0.010}^{+0.007}$ \\
(10935,12020)\footnotemark[2] & 2009.84 & $2.080_{-0.008}^{+0.009}$ \\
(10936,13177)\footnotemark[2] & 2010.83 & $2.070_{-0.009}^{+0.009}$ \\
14229 & 2012.37 & $2.050_{-0.008}^{+0.009}$ \\
14480 & 2013.38 & $2.075_{-0.009}^{+0.009}$ \\
14481 & 2014.36 & $2.045_{-0.009}^{+0.009}$
\end{tabular}
\footnotetext[1]{Errors are $1\sigma$.}
\footnotetext[2]{The two ObsIDs, which were taken close together in
time with the same instrument setup, are merged prior to spectral analysis.}
\end{ruledtabular}
\end{table}
%-------------------------------------------

Since the Cas~A NS belongs to a class of NSs known as central compact objects
(CCOs) and three members of this class have surface magnetic fields
$\sim 10^{10}-10^{11}\mbox{ G}$
(the interior field may be much higher; see \cite{halperngotthelf10,ho13}),
we also attempt to fit the relatively low magnetic field hydrogen atmosphere
model spectra described in \cite{ho13};
note that the model spectra currently available at field strengths
$(1,4,7,10)\times 10^{10}\mbox{ G}$ are computed for only surface gravity
$=2.4\times10^{14}\mbox{ cm s$^{-2}$}$.
At the high temperatures present at early NS ages, nuclear burning rapidly
removes surface hydrogen and helium \cite{changbildsten04,changetal10}.
However, non-hydrogen atmosphere models for the relevant magnetic fields do
not currently exist.
Also, even though the hydrogen model spectra we use are for a fully
ionized atmosphere, the fitted temperatures are high ($\Teff>10^6\mbox{ K}$),
such that spectral features due to any trace amounts of bound species do not
significantly affect the spectra \cite{potekhinetal14}.
The resulting fits can be good (with $\chisqr\approx 1$ for 337 degrees of
freedom) but have unrealistically small NS mass and radius
($<0.4\,\Msun$ and $\sim 5\mbox{ km}$), and thus we do not consider these
models further.

\section{Neutron star model \label{sec:model}}

\subsection{Equation of state \label{sec:eos}}
To construct non-rotating equilibrium NSs, we solve the
Tolman-Oppenheimer-Volkoff relativistic equations of stellar structure
(see, e.g., \cite{shapiroteukolsky83}), supplemented by the EOS.
We consider three nuclear EOSs:
The first is APR, specifically A18+$\delta v$+UIX$^\ast$ \cite{akmaletal98},
with the neutron and proton effective masses given by the analytic formula
in \cite{pageetal04},
and is the same EOS that is used in \cite{pageetal04,pageetal09,pageetal11}.
The other two are BSk20 and BSk21 \cite{fantinaetal13}, which are calculated
using the analytic functions in \cite{potekhinetal13},
with the nucleon effective masses given by the analytic formula in
\cite{chameletal09} and parameters in \cite{gorielyetal10}.
BSk20 and BSk21 use generalized Skyrme forces and are constructed to satisfy
various experimental constraints (see \cite{potekhinetal13} and references
therein)
and to be similar to APR of \cite{akmaletal98} and V18 of \cite{lischulze08},
respectively.
In addition, the crust composition predicted by BSk21 is compatible with
the recent nuclear mass measurement of \cite{wolfetal13}.
All three EOSs produce a NS with maximum mass $>2\,\Msun$, as needed to
match the (highest) observed NS masses of $1.97\pm 0.04\,\Msun$
\cite{demorestetal10} and $2.01\pm 0.04\,\Msun$ \cite{antoniadisetal13}.

\subsection{Thermal evolution \label{sec:cool}}

The evolution of the interior temperature of an isolated NS is determined by
the relativistic equations of energy balance and heat flux
(see, e.g., \cite{shapiroteukolsky83,yakovlevpethick04}).
We use the NS cooling code described in \cite{gnedinetal01}.
The revised Cooper pairing emissivity from \cite{pageetal09} is included.
NS models with $M>\Mdu$ undergo (fast) direct Urca cooling, and
$\Mdu=1.96\,\Msun$ for APR and $\Mdu=1.59\,\Msun$ for BSk21, while
BSk20 does not produce NSs that undergo direct Urca cooling for any mass.
Note that there are a few very cold NSs in binary systems, such as
SAX~J1808.4$-$3658, which suggest direct Urca cooling should occur for
some NS masses \cite{heinkeetal09}.
The initial temperature is taken to be a constant $Te^\Phi=10^{10}\mbox{ K}$,
where $\Phi$ is the metric function which corresponds to the gravitational
potential in the Newtonian limit \cite{shapiroteukolsky83}.

The outer layers (envelope) of the NS crust serve as a heat blanket, and
there can exist a large temperature gradient between the bottom of the
envelope (at $\rho\sim 10^{10}\mbox{ g cm$^{-3}$}$) and the surface
\cite{gudmundssonetal82,yakovlevpethick04}.
Light elements have higher thermal conductivity and make the envelope more
heat transparent \cite{potekhinetal03},
while high temperatures of young NSs cause rapid nuclear burning and
removal of surface hydrogen and helium \cite{changbildsten04,changetal10}.
Therefore we consider several cases.  One is when the amount of carbon that
covers the NS is very small ($\Delta M\sim 10^{-18}\Msun$) and is only
sufficient to produce an optically thick atmosphere of carbon, which is
needed to fit the X-ray spectra of the Cas~A NS \cite{hoheinke09}.
The envelope beneath this atmosphere is then composed of iron, and we use
the relation between the surface and envelope temperature from
\cite{potekhinetal03}.
The other cases are when there is a carbon layer
(with carbon $\Delta M\sim 10^{-15}$, $10^{-11}$, or $10^{-8}\Msun$)
that extends down from the atmosphere to the bottom of the NS envelope.

\subsection{Superfluid and superconducting gap models \label{sec:sf}}

Superfluidity and superconductivity have two important effects on neutrino
emission and NS cooling:
(1) suppression of heat capacities and emission mechanisms, like modified
Urca processes, that involve superfluid and superconducting constituents and
(2) enhanced emission due to Cooper pairing of nucleons when the
temperature decreases just below the critical value
(see \cite{yakovlevpethick04,pageetal06}, for reviews).
These two effects on the temperature evolution will be shown below
in Secs.~\ref{sec:ns}-\ref{sec:nt}.

The critical temperatures for superfluidity are approximately related
to the superfluid energy gap $\Delta$ by
$k\Tc\approx 0.5669\Delta$ for the singlet (isotropic pairing) gap and
\be
k\Tc = 0.5669\frac{\Delta}{2^{1/2}\Gamma_0}=0.1187\Delta
 \sim 0.5669\frac{\Delta}{\sqrt{8\pi}},
\ee
where $\ln\Gamma_0\approx 1.22$, for the triplet (anisotropic pairing) gap
\cite{takatsukatamagaki71,amundsenostgaard85b,baldoetal92}.
Furthermore, what is required for NS cooling calculations is the critical
temperature as a function of mass or baryon density, $\Tc(\rho)$ or $\Tc(\nb)$,
respectively.
To convert gap energy as a function of Fermi momentum $\Delta(\kf)$ into
$\Tc(\rho)$, where $\hbar\kf=\hbar(3\pi^2n_{\mathrm{x}})^{1/3}$ and
$n_{\mathrm{x}}$ are the Fermi momentum and number density, respectively,
for particle species ${\mathrm{x}}$, an EOS must be used.  Below we give
examples of $\Tc(\rho)$ and $\Tc(r/R)$ using the APR, BSk20, and BSk21 EOSs.  

\begin{figure}
\resizebox{!}{0.30\textheight}{\includegraphics{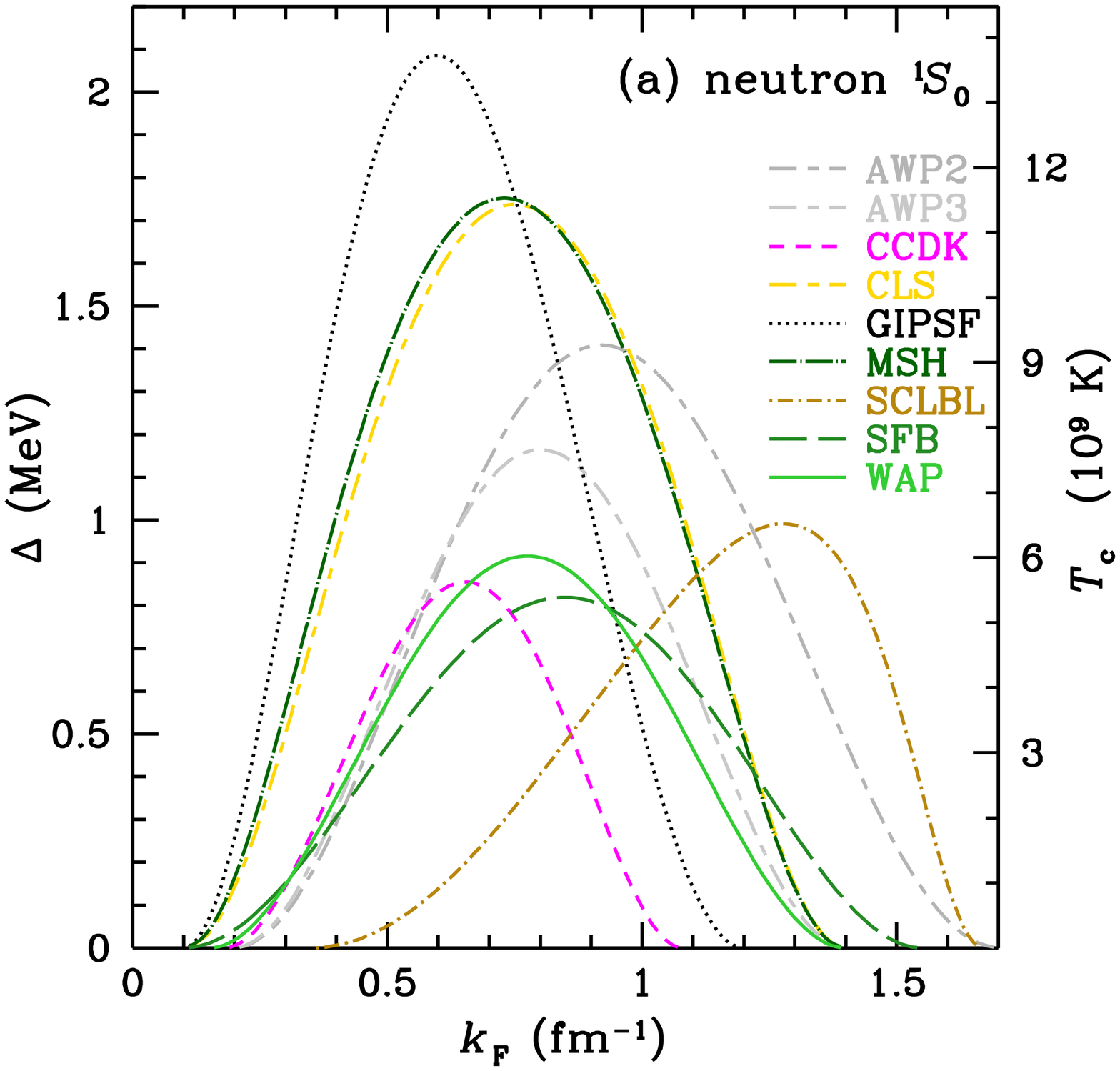}}
\resizebox{!}{0.30\textheight}{\includegraphics{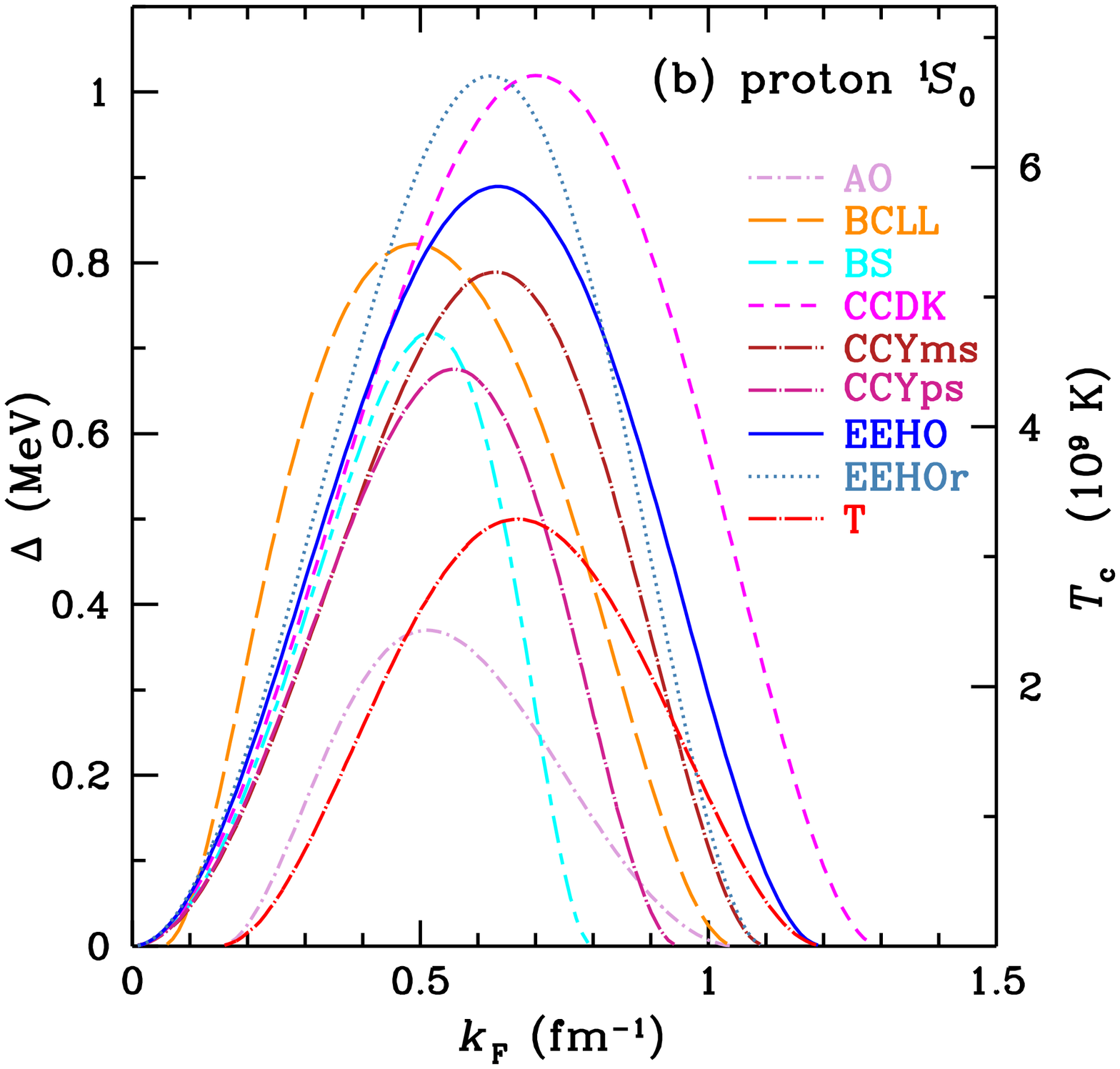}}
\resizebox{!}{0.30\textheight}{\includegraphics{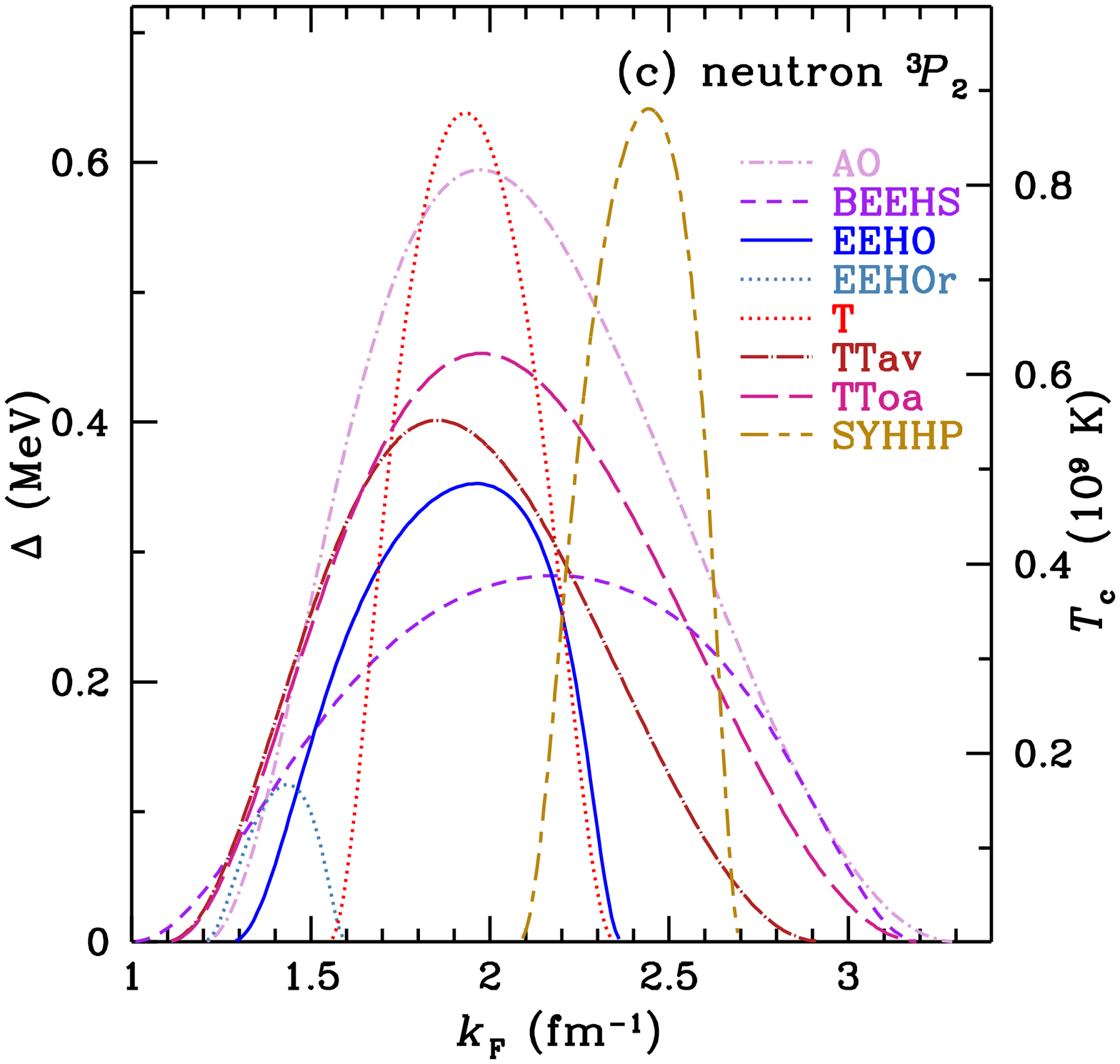}}
% \resizebox{\hsize}{!}{\includegraphics{sfkns.eps} \includegraphics{sfkps.eps} \includegraphics{sfknt.eps}}
\caption{
(Color online)
Top: Neutron singlet gap energy (left axis) and critical temperature
(right axis).
Middle: Proton singlet gap energy and critical temperature.
Bottom: Neutron triplet gap energy and critical temperature.
Labels indicate particular gap models (see Table~\ref{tab:sfgap}).
\label{fig:sfk}}
\end{figure}

We use the parameterization for the gap energy similar to that used by
\cite{kaminkeretal01,lombardoschulze01,kaminkeretal02,anderssonetal05}
\be
\Delta(\kf) = \Delta_0\frac{(\kf-k_0)^2}{(\kf-k_0)^2+k_1}
 \frac{(\kf-k_2)^2}{(\kf-k_2)^2+k_3}, \label{eq:sfparam}
\ee
where $\Delta_0$, $k_0$, $k_1$, $k_2$, and $k_3$ are fit parameters.
We determine these fit parameters for various superfluid gap models
from the literature, and the values are given in Table~\ref{tab:sfgap}.
Figure~\ref{fig:sfk} shows the gap models.
We note another model for neutron singlet is that of
\cite{gezerliscarlson08,gezerliscarlson10};
however their results are only at three (low) values of $\kfn$ and appear
similar to the CLS and MSH models when extrapolated to higher $\kfn$.

%-------------------------------------------
\begin{table}
\caption{Superfluid gap parameters. \label{tab:sfgap}}
\begin{ruledtabular}
\begin{tabular}{ccccccc}
Gap & $\Delta_0$ & $k_0$ & $k_1$ & $k_2$ & $k_3$ & Ref. \\
model  & (MeV) & (fm$^{-1}$) & (fm$^{-2}$) & (fm$^{-1}$) & (fm$^{-2}$) & \\
\hline
 \multicolumn{7}{c}{Neutron singlet (ns)} \\
AWP2 & 28 & 0.20 & 1.5 & 1.7 & 2.5 & \cite{ainsworthetal89} \\
AWP3 & 50 & 0.20 & 2.0 & 1.4 & 2.0 & \cite{ainsworthetal89} \\
CCDK & 127 & 0.18 & 4.5 & 1.08 & 1.1 & \cite{chenetal93} \\
CLS & 2.2 & 0.18 & 0.06 & 1.3 & 0.03 & \cite{caoetal06,gandolfietal09} \\
GIPSF & 8.8 & 0.18 & 0.1 & 1.2 & 0.6 & \cite{gandolfietal09,gandolfietal08} \\
MSH & 2.45 & 0.18 & 0.05 & 1.4 & 0.1 & \cite{margueronetal08,gandolfietal09} \\
SCLBL & 4.1 & 0.35 & 1.7 & 1.67 & 0.06 & \cite{schulzeetal96} \\
SFB & 45 & 0.10 & 4.5 & 1.55 & 2.5 & \cite{schwenketal03} \\
WAP & 69 & 0.15 & 3.0 & 1.4 & 3.0 & \cite{wambachetal93,schwenketal03} \\
\multicolumn{7}{c}{Proton singlet (ps)} \\
AO & 14 & 0.15 & 0.22 & 1.05 & 3.8 & \cite{amundsenostgaard85,elgaroyetal96a} \\
BCLL & 1.69 & 0.05 & 0.07 & 1.05 & 0.16 & \cite{baldoetal92,elgaroyetal96a} \\
BS & 17 & 0.0 & 2.9 & 0.8 & 0.08 & \cite{baldoschulze07} \\
CCDK & 102 & 0.0 & 9.0 & 1.3 & 1.5 & \cite{chenetal93,elgaroyetal96a} \\
CCYms & 35 & 0.0 & 5.0 & 1.1 & 0.5 & \cite{chaoetal72} \\
CCYps & 34 & 0.0 & 5.0 & 0.95 & 0.3 & \cite{chaoetal72} \\
EEHO & 4.5 & 0.0 & 0.57 & 1.2 & 0.35 & \cite{elgaroyetal96a} \\
EEHOr\footnote{Fit parameters given by model $e$ of \cite{anderssonetal05}.}
 & 61 & 0.0 & 6.0 & 1.1 & 0.6 & \cite{elgaroyetal96c} \\
T & 48 & 0.15 & 2.1 & 1.2 & 2.8 & \cite{takatsuka73} \\
\multicolumn{7}{c}{Neutron triplet (nt)} \\
AO & 4.0 & 1.2 & 0.45 & 3.3 & 5.0 & \cite{amundsenostgaard85b} \\
BEEHS\footnote{Fit to the BHF spectra from Fig.~4 of \cite{baldoetal98},
not BHFm$^\ast$, since \cite{baldoetal98} state that an effective mass
approximation should not be used when calculating the gap.}
 & 0.45 & 1.0 & 0.40 & 3.2 & 0.25 & \cite{baldoetal98} \\
EEHO\footnote{Fit parameters given by model $l$ of \cite{anderssonetal05}.}
 & 0.48 & 1.28 & 0.1 & 2.37 & 0.02 & \cite{elgaroyetal96b} \\
EEHOr & 0.23 & 1.2 & 0.026 & 1.6 & 0.0080 & \cite{elgaroyetal96c} \\
SYHHP\footnote{Replaces the deep model given in \cite{hoetal12}.}
 & 1.0 & 2.08 & 0.04 & 2.7 & 0.013 & \cite{shterninetal11} \\
T & 1.2 & 1.55 & 0.05 & 2.35 & 0.07 & \cite{takatsuka72,amundsenostgaard85b} \\
TTav & 3.0 & 1.1 & 0.60 & 2.92 & 3.0 & \cite{takatsukatamagaki04} \\
TToa & 2.1 & 1.1 & 0.60 & 3.2 & 2.4 & \cite{takatsukatamagaki04} \\
\end{tabular}
\end{ruledtabular}
\end{table}
%-------------------------------------------

%-------------------------------------------
%-------------------------------------------

\section{Results \label{sec:results}}

\subsection{Mass and radius from \Chandra X-ray spectra \label{sec:fitsp}}

\begin{figure}
\resizebox{0.95\hsize}{!}{\includegraphics{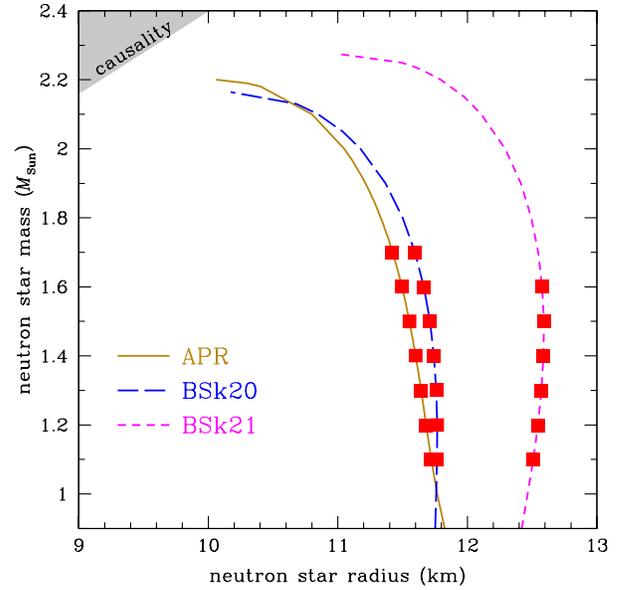}}
\caption{
(Color online)
Neutron star mass versus radius for three nuclear EOSs: APR (solid),
BSk20 (long-dashed), and BSk21 (short-dashed).
Squares indicate ($M$,$R$)-values which produce good fits to \Chandra ACIS-S
Graded data at a 90\% confidence level.
\label{fig:fitspmr}}
\end{figure}

\begin{figure}
\resizebox{0.95\hsize}{!}{\includegraphics{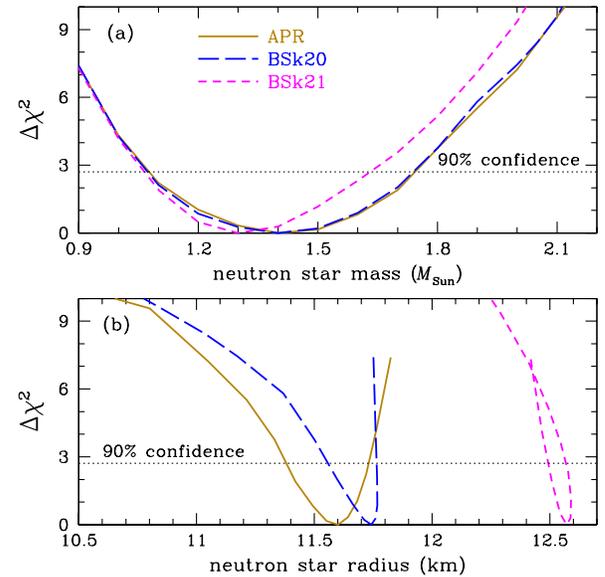}}
\caption{
(Color online)
Best-fit to \Chandra ACIS-S Graded data, as determined by $\Delta\chi^2$
as a function of NS mass (top) and radius (bottom) for three nuclear EOSs:
APR (solid), BSk20 (long-dashed), and BSk21 (short-dashed).
Dotted lines indicate the 90\% confidence level.
\label{fig:fitsp}}
\end{figure}

Figures~\ref{fig:fitspmr} and \ref{fig:fitsp} show the results of our
simultaneous fit to all nine sets of \Chandra ACIS-S Graded spectra
(see Table~\ref{tab:tempgraded}).
Here we fit for the grade migration parameter (one for observations with a
3.04~s frame time and another for observations with a 3.24~s frame time;
see \cite{heinkeho10,elshamoutyetal13} for details),
hydrogen column density, and surface temperature $\Teff$
but hold each at a single value for all observations, except $\Teff$.
We also hold mass $M$ and radius $R$ to a single value, but rather than
allow them to take on any value in their respective
parameter space, we only use pairs of values ($M$,$R$) that are produced by
each EOS considered herein.  Thus $M$-$R$ confidence contours collapse down to
confidence levels along an $M$-$R$ sequence for each EOS; this is shown in
Fig.~\ref{fig:fitspmr}.
We see from Fig.~\ref{fig:fitsp} that the best-fit NS mass at $\approx 90\%$
confidence is $M\approx 1.4\pm0.3\,\Msun$ for any of the three EOSs.
Meanwhile the best-fit NS radius at $\approx 90\%$ confidence is
$R\approx 11.6_{-0.2}^{+0.1}\mbox{ km}$ for APR,
$11.7\pm 0.1\mbox{ km}$ for BSk20, and $12.55\pm0.05\mbox{ km}$ for BSk21.
The peculiar shape of the fit for $R$ for BSk21 is due to the nearly constant
NS radius predicted by this EOS for $M\approx 1.1-1.8\,\Msun$.
Finally we note that the grade migration parameter is $\approx 0.2-0.35$ and
hydrogen column density is $\approx (1.6-1.8)\times 10^{22}\mbox{ cm$^{-2}$}$
(see also \cite{heinkeho10,elshamoutyetal13}),
both of which are proportional to the assumed value of $M$.
Since regions of the supernova remnant near the NS have hydrogen column
density $\approx (1.7-2)\times 10^{22}\mbox{ cm$^{-2}$}$
\cite{hwanglaming12}, a higher NS mass ($M\gtrsim 1.6\,\Msun$) is favored.

%-------------------------------------------
%-------------------------------------------

%-------------------------------------------
%-------------------------------------------

\subsection{Neutron crust superfluid \label{sec:ns}}

\begin{figure}
\resizebox{0.95\hsize}{!}{\includegraphics{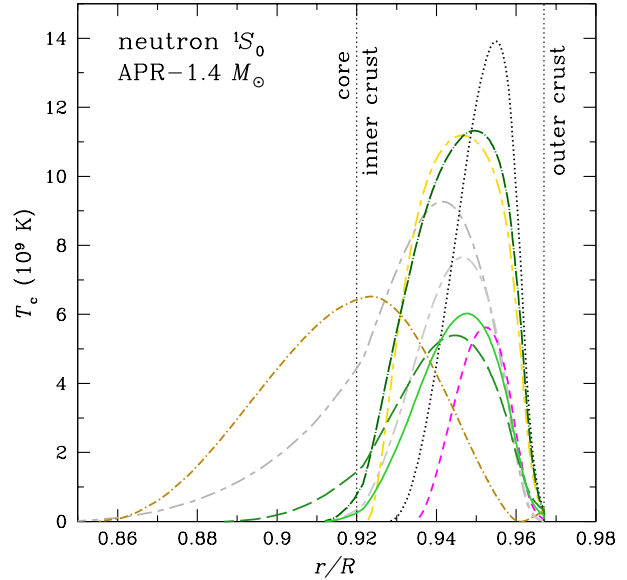}}
\caption{
(Color online)
Critical temperature $\Tc$ for neutron singlet superfluidity as a function of
fractional radius of a NS constructed using the APR EOS
($M=1.4\,\Msun$, $R=11.6\mbox{ km}$).
Different curves correspond to different gap models that are shown in
Fig.~\ref{fig:sfk}.
Vertical dotted lines denote the boundaries between the core, inner crust,
and outer crust of the NS.
\label{fig:tcrns_apr}}
\end{figure}

We first consider only the introduction of the neutron singlet gap into the
cooling simulations, and we only display results using the APR EOS for
simplicity.
Figure~\ref{fig:tcrns_apr} shows the critical temperature $\Tc$ for the
onset of neutron superfluidity in the singlet state as a function of relative
radius $r/R$.
Most neutron singlet gap models are primarily confined to the inner crust.
However, we see that a few (i.e., AWP2, SCLBL, and SFB) extend into the core.

%-------------------------------------------
%-------------------------------------------

\begin{figure}
\resizebox{0.95\hsize}{!}{\includegraphics{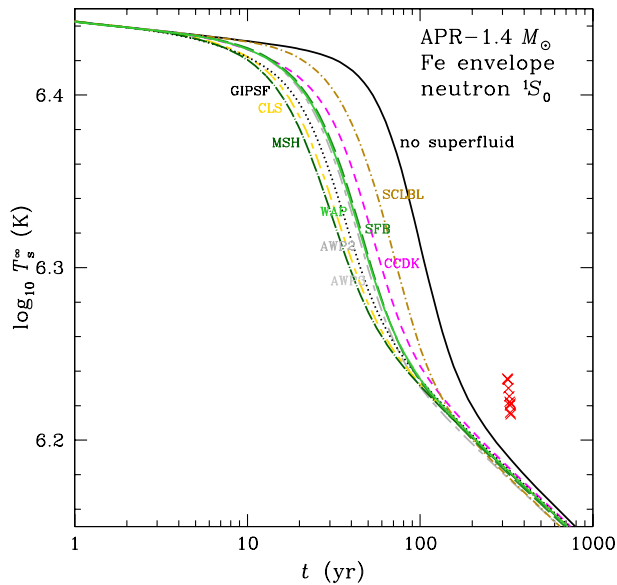}}
\caption{
(Color online)
Redshifted surface temperature $\Tsinfty$ as a function of age for a
$1.4\,\Msun$ APR NS with an iron envelope.
Different curves are cooling simulations using one corresponding neutron
singlet gap model (see Fig.~\ref{fig:sfk}),
while the curve labeled ``no superfluid'' is a simulation that does not
include any superfluid components.
Crosses are the observed temperatures of the Cas~A NS.
\label{fig:cool_ns}}
\end{figure}

Figure~\ref{fig:cool_ns} shows the time evolution of the redshifted surface
temperature $\Tsinfty=\Ts/(1+\zg)$, where $1+\zg=(1-2GM/c^2R)^{-1/2}$ is
the gravitational redshift.
The temperature evolution (or cooling curve) labeled ``no superfluid'' is
calculated using a $1.4\,\Msun$ NS with the APR EOS, iron envelope, and
no superfluid or superconducting gap models.
The other cooling curves are calculated using the same NS model
but including one neutron singlet gap model (denoted by the labels; see
Table~\ref{tab:sfgap}).
As mentioned in Sec.~\ref{sec:sf}, the two primary effects of
superfluidity/superconductivity on NS cooling are suppression of neutrino
emission processes that involve particles that are superfluid or
superconducting and enhancement of cooling due to neutrino production
during Cooper pairing.
Here we see that the second effect (more rapid cooling) is dominant in the
case of the onset of neutron superfluidity in the singlet state
(as well as suppression of the neutron heat capacity, which is also included
here; see also \cite{pageetal09}).
All neutron singlet gap models produce cooling curves that show a rapid
temperature decline at an earlier age than the cooling curve generated
without including superfluidity; similar results are seen in \cite{pageetal09}.
Note that the general behavior of rapid decline is due to thermal
relaxation of the NS.
At very early times, the NS core cools more rapidly than the crust via
the stronger neutrino emission that occurs in the core, so that the crust
is generally at higher temperatures.
A cooling wave travels from the core to the surface, bringing the NS to
a relaxed, isothermal state.
The relaxation time is $\sim 10-100\mbox{ yr}$, depending on the properties
of the crust \cite{lattimeretal94,gnedinetal01,yakovlevetal11}.
Incidentally,
formation of the inner and outer crusts begins at $\sim 1\mbox{ hr}$ and
$\sim 1\mbox{ day}$, respectively, and is mostly complete after
$\sim 1\mbox{ month}$ and $\sim 1\mbox{ yr}$, respectively
\cite{aguileraetal08,hoetal12}.
For a much lower NS mass or thicker crust, thermal relaxation may require a
few hundred years.
Nevertheless we see that thermal relaxation, as well as the effects of any
of the neutron singlet gap models, occurs well before the time of our
observations of the Cas~A NS.  Therefore Cas~A is not useful for
constraining the epoch of thermal relaxation or these gap models
(cf. \cite{blaschkeetal12}).

%-------------------------------------------
%-------------------------------------------

\subsection{Proton core superconductor \label{sec:ps}}

\begin{figure}
\resizebox{0.95\hsize}{!}{\includegraphics{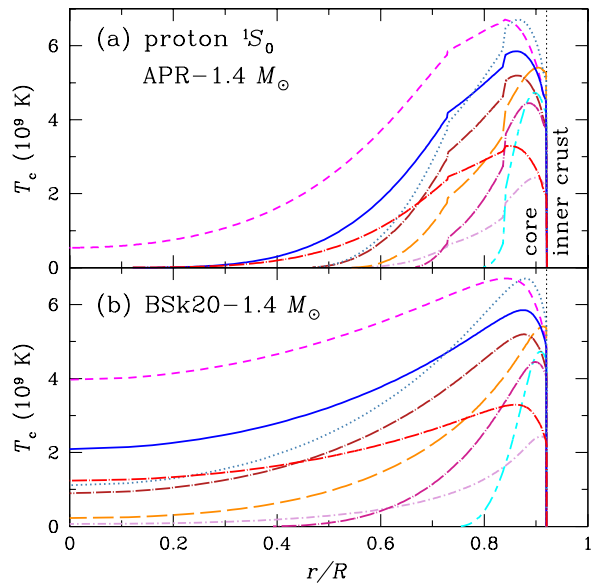}}
\caption{
(Color online)
Critical temperature $\Tc$ for proton superconductivity as a function of
fractional radius of a NS constructed using the
APR EOS ($M=1.4\,\Msun$, $R=11.6\mbox{ km}$; top panel)
and BSk20 EOS ($M=1.4\,\Msun$, $R=11.7\mbox{ km}$; bottom panel).
Different curves correspond to different proton singlet gap models that
are shown in Fig.~\ref{fig:sfk}.
Vertical dotted line denotes the boundary between the core and inner crust
of the NS.
\label{fig:tcrps}}
\end{figure}

We now consider (only) the introduction of the proton singlet gap into the
cooling simulations.
Figure~\ref{fig:tcrps} shows the critical temperature $\Tc$ for the
onset of proton superconductivity in the NS core as a function of relative
radius $r/R$ for the APR and BSk20 EOSs.
For most gap models using the APR EOS and high temperatures
($T>10^8\mbox{ K}$), protons in the superconducting state only occupy a
fractional radius of 0.1--0.3 for a $1.4\,\Msun$ NS.  Only the CCDK gap
model can produce a NS that has a completely superconducting core of protons.
On the other hand, we see that proton superconductivity
can extend throughout the core for most gap models using the BSk20 EOS.
This difference between the two EOSs is due to the larger proton fraction
(at the same baryon density) in APR compared to BSk20.
The critical temperature (or gap energy) increases, reaches a maximum, and
then decreases as a function of Fermi momentum $\kfp$ or proton density
$n_{\mathrm p}$ (see Fig.~\ref{fig:sfk}).
The larger proton fraction for APR means that we can see to larger $\kfp$
where the gap energy tail becomes small.
The proton superconductor critical temperatures for the BSk21 EOS are
intermediate between the ones for APR and BSk20.

%-------------------------------------------
%-------------------------------------------

\begin{figure}
\resizebox{0.95\hsize}{!}{\includegraphics{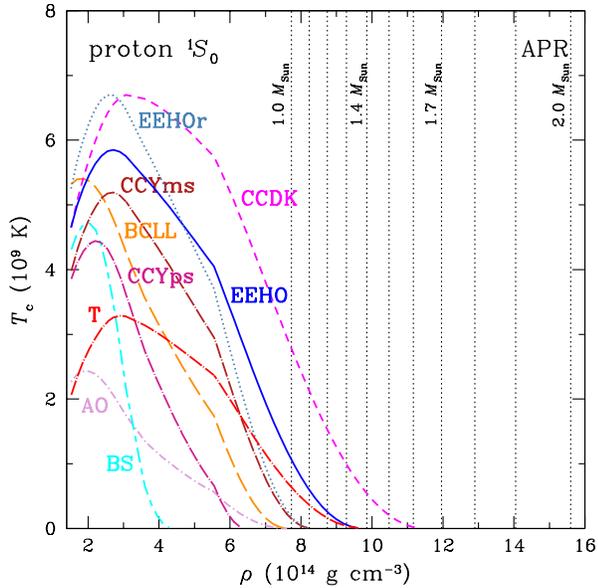}}
\caption{
(Color online)
Critical temperature $\Tc$ for proton superfluidity as a function of
mass density of a NS constructed using the APR EOS.
Different curves correspond to different proton singlet gap models that
are shown in Fig.~\ref{fig:sfk}.
Vertical dotted lines denote the core density of NSs of different mass.
\label{fig:tcdps_apr}}
\end{figure}

Figure~\ref{fig:tcdps_apr} shows the critical temperature as a function of
density.  Also shown by the vertical lines is the central density of an
APR NS of various masses.
Only for the strong CCDK gap model does proton superconductivity extend
down into the center of NSs with $M>1.3\,\Msun$.
In subsequent sections, we will consider only the CCDK model for
the proton superconducting gap energy.

%-------------------------------------------
%-------------------------------------------

\begin{figure}
\resizebox{0.95\hsize}{!}{\includegraphics{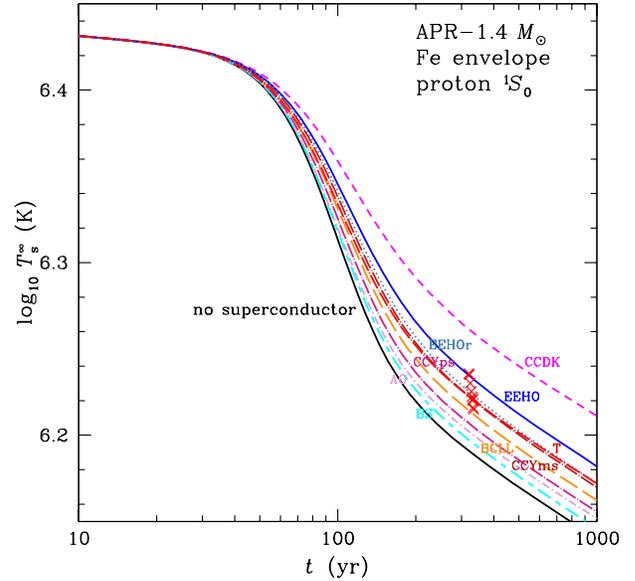}}
\caption{
(Color online)
Redshifted surface temperature $\Tsinfty$ as a function of age for a
$1.4\,\Msun$ APR NS with an iron envelope.
Different curves are cooling simulations using one corresponding proton
gap model (see Fig.~\ref{fig:sfk}),
while the curve labeled ``no superconductor'' is a simulation that does not
include a superconductor component.
Crosses are the observed temperatures of the Cas~A NS.
\label{fig:cool_ps}}
\end{figure}

Figure~\ref{fig:cool_ps} shows cooling curves calculated using a $1.4\,\Msun$
NS with the APR EOS and iron envelope and including one proton singlet gap
model (denoted by the labels; see Table~\ref{tab:sfgap}).
The cooling curve labeled ``no superconductor'' is calculated with no superfluid
or superconducting gap models.
As a result of low proton fractions,
we see that the first effect (less efficient cooling) discussed in
Sec.~\ref{sec:sf}, i.e., suppression of neutrino emission processes that
involve protons, is dominant in the case of the onset of proton
superconductivity.
For the BSk20 and BSk21 EOSs, the proton superconductor critical temperatures
extend to greater fractions of the NS core (see Fig.~\ref{fig:tcrps}), and
as a result, this suppression will be stronger and will produce more rapid
temperature drops when the core {\it neutrons} become superfluid and emit
Cooper-pairing neutrinos.

%-------------------------------------------
%-------------------------------------------

\subsection{Neutron core superfluid \label{sec:nt}}

\begin{figure}
\resizebox{0.95\hsize}{!}{\includegraphics{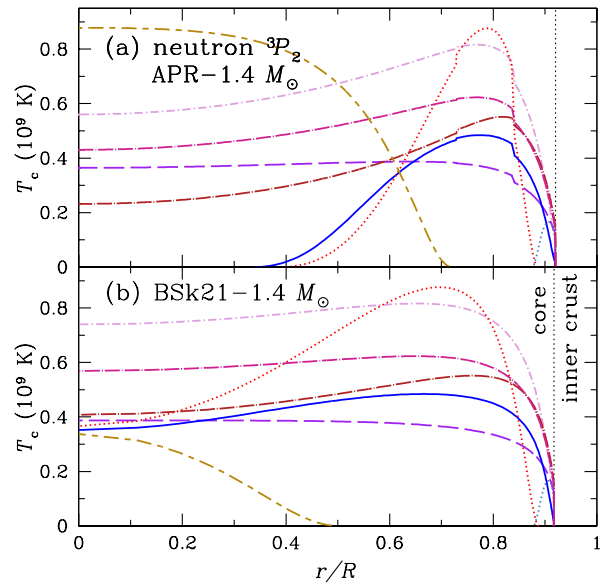}}
\caption{
(Color online)
Critical temperature $\Tc$ for neutron triplet superfluidity as a function
of fractional radius of a NS constructed using the
APR EOS ($M=1.4\,\Msun$, $R=11.6\mbox{ km}$; top panel)
and BSk21 EOS ($M=1.4\,\Msun$, $R=12.6\mbox{ km}$; bottom panel).
Different curves correspond to different gap models that are shown in
Fig.~\ref{fig:sfk}.
Vertical dotted line denotes the boundary between the core and inner crust
of the NS.
\label{fig:tcrnt}}
\end{figure}

\begin{figure}
\resizebox{0.95\hsize}{!}{\includegraphics{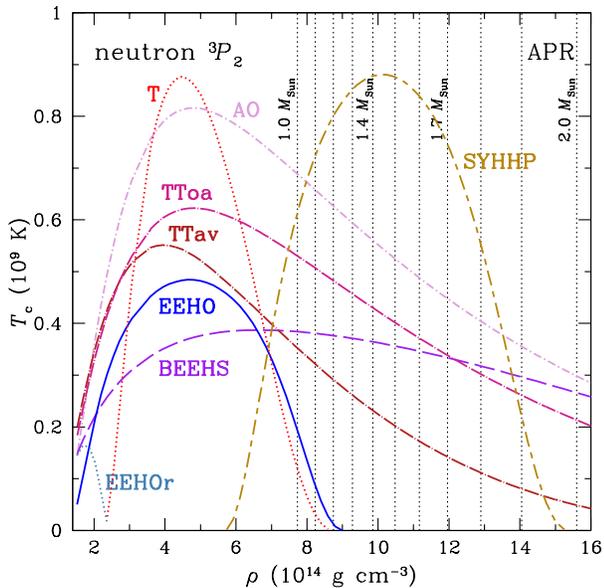}}
\caption{
(Color online)
Critical temperature $\Tc$ for neutron triplet superfluidity as a function
of mass density of a NS constructed using the APR EOS.
Different curves correspond to different gap models that are shown in
Fig.~\ref{fig:sfk}.
Vertical dotted lines denote the core density of NSs of different mass.
\label{fig:tcdnt_apr}}
\end{figure}

Finally we consider the neutron triplet gap.
Figure~\ref{fig:tcrnt} shows the critical temperature $\Tc$ for the
onset of neutron superfluidity in the triplet state in the NS core as a
function of relative radius $r/R$ using the APR and BSk21 EOSs.
Unlike proton superconductivity, strong neutron superfluidity can extend
throughout the core for many triplet gap models.
This is particularly the case for the BSk21 EOS (BSk20 is more similar to
APR); thus a much larger fraction of the NS can become superfluid with
the BSk21 EOS, except for the SYHHP gap model.
Figure~\ref{fig:tcdnt_apr} shows the critical temperature as a function of
density, as well as the central density of an APR NS of various masses.
It is clear that the entire core of all NS masses can be in a superfluid
state.
We note here the dramatically different behavior of the SYHHP gap model
compared to all other models.
This is because, unlike the other gap models which are derived from nuclear
theory calculations, SYHHP is a phenomenological model constructed to fit
the observed cooling behavior of NSs \cite{gusakovetal04}.

%-------------------------------------------
%-------------------------------------------

%-------------------------------------------
%-------------------------------------------

\begin{figure}
\resizebox{0.95\hsize}{!}{\includegraphics{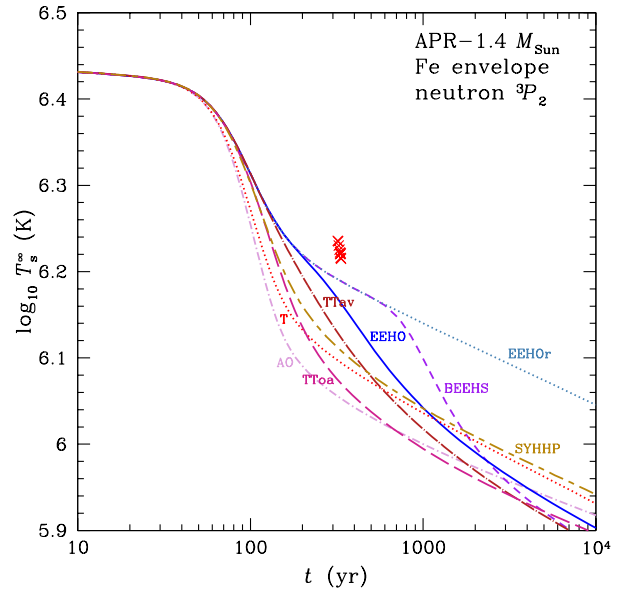}}
\caption{
(Color online)
Redshifted surface temperature $\Tsinfty$ as a function of age for a
$1.4\,\Msun$ APR NS with an iron envelope.
Different curves are cooling simulations using one corresponding neutron
triplet gap model (see Fig.~\ref{fig:sfk}),
while the curve labeled ``EEHOr'' is a simulation that effectively does not
include any superfluid components since this gap model has no effect for
the relevant ages shown.
Crosses are the observed temperatures of the Cas~A NS.
\label{fig:cool_nt}}
\end{figure}

Figure~\ref{fig:cool_nt} shows cooling curves calculated using a $1.4\,\Msun$
NS with the APR EOS and iron envelope and including one neutron triplet gap
model (denoted by the labels; see Table~\ref{tab:sfgap}).
The EEHOr gap model has very low critical temperatures and occupies a very
small fraction of the NS (see Fig.~\ref{fig:tcrnt});
therefore the cooling simulation which uses this gap model is effectively one
without any core superfluid for the ages ($<10^4\mbox{ yr}$) considered here.

%-------------------------------------------
%-------------------------------------------

\begin{figure*}
\resizebox{\hsize}{!}{\includegraphics{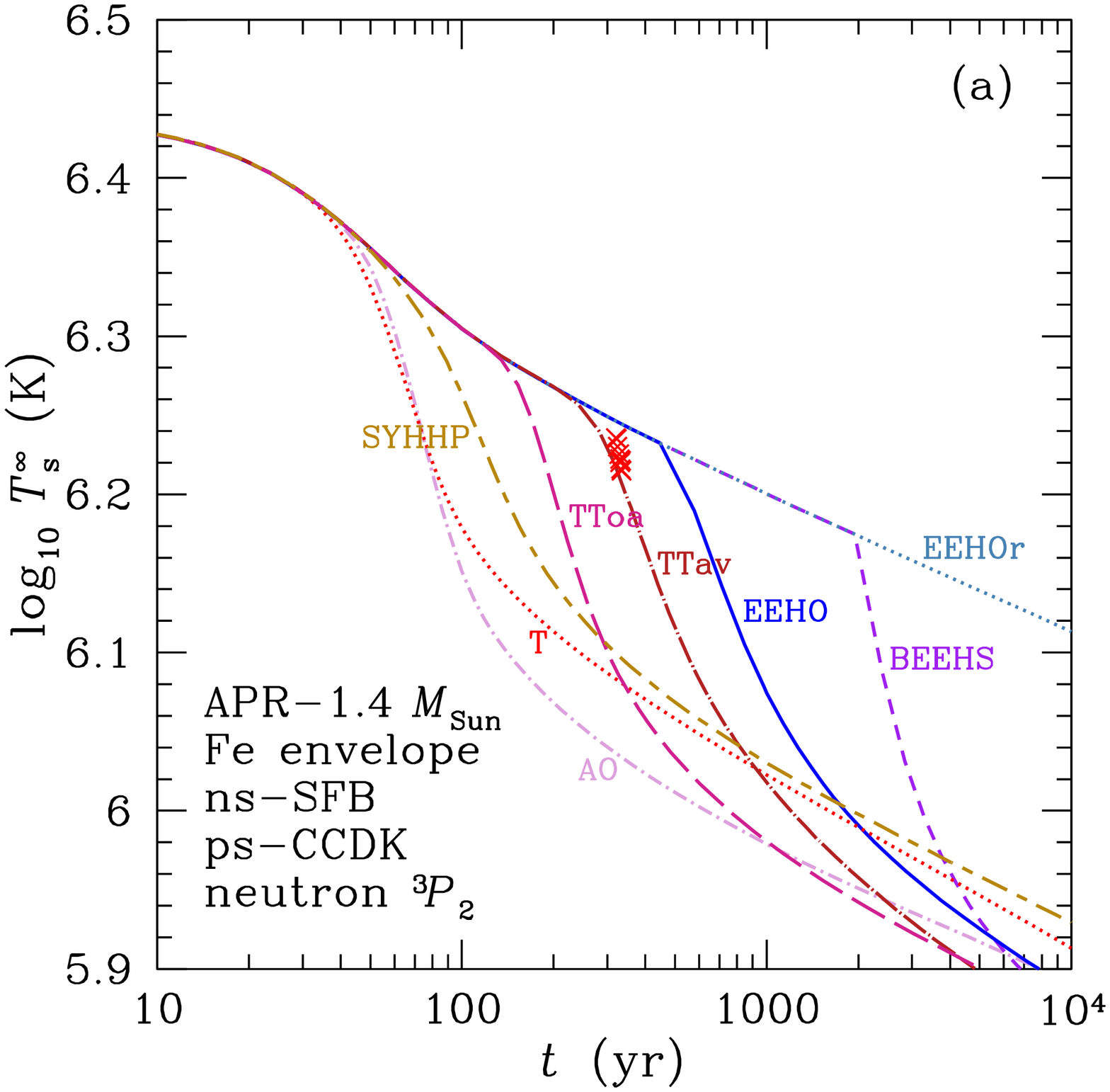} \includegraphics{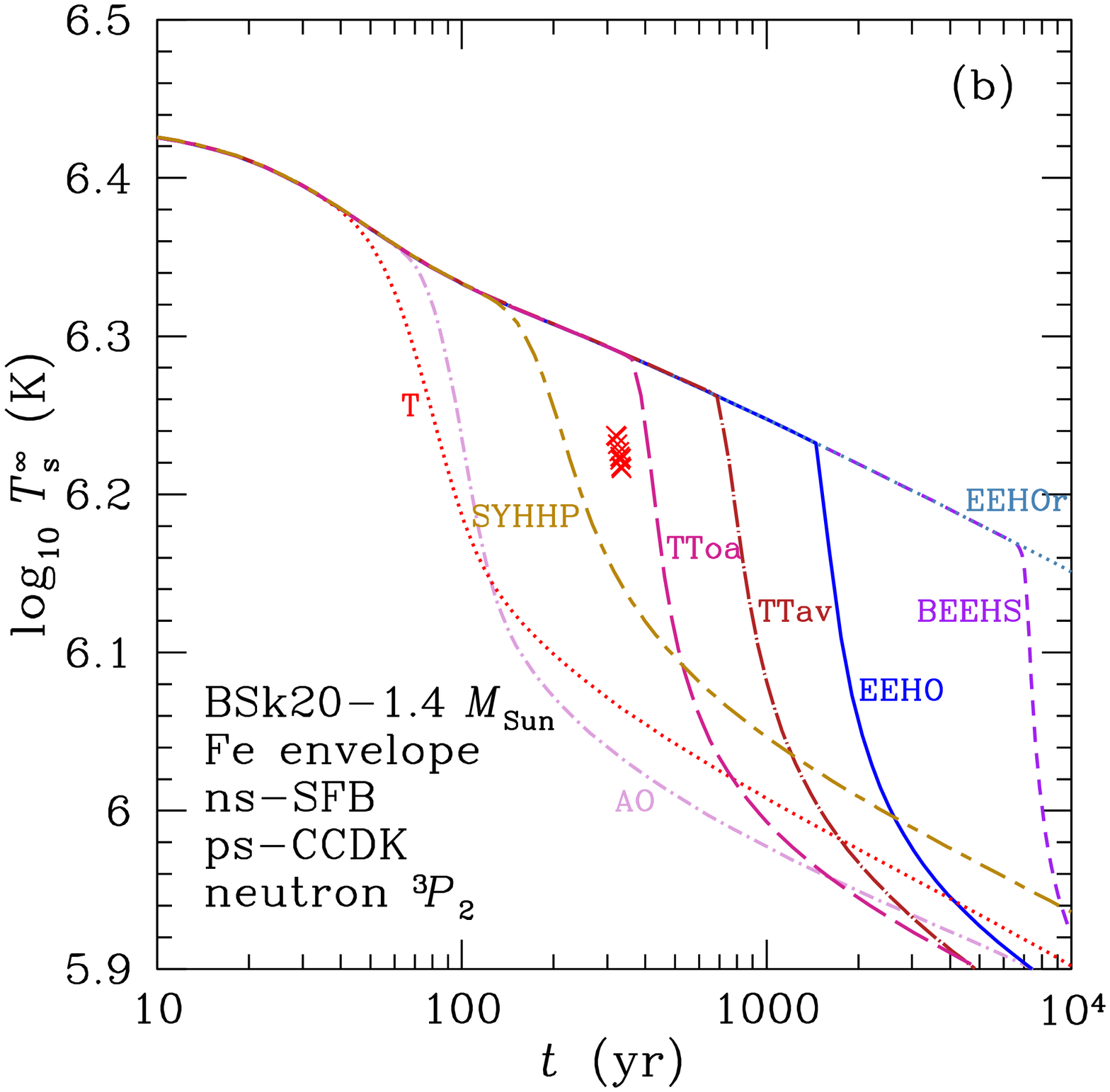} \includegraphics{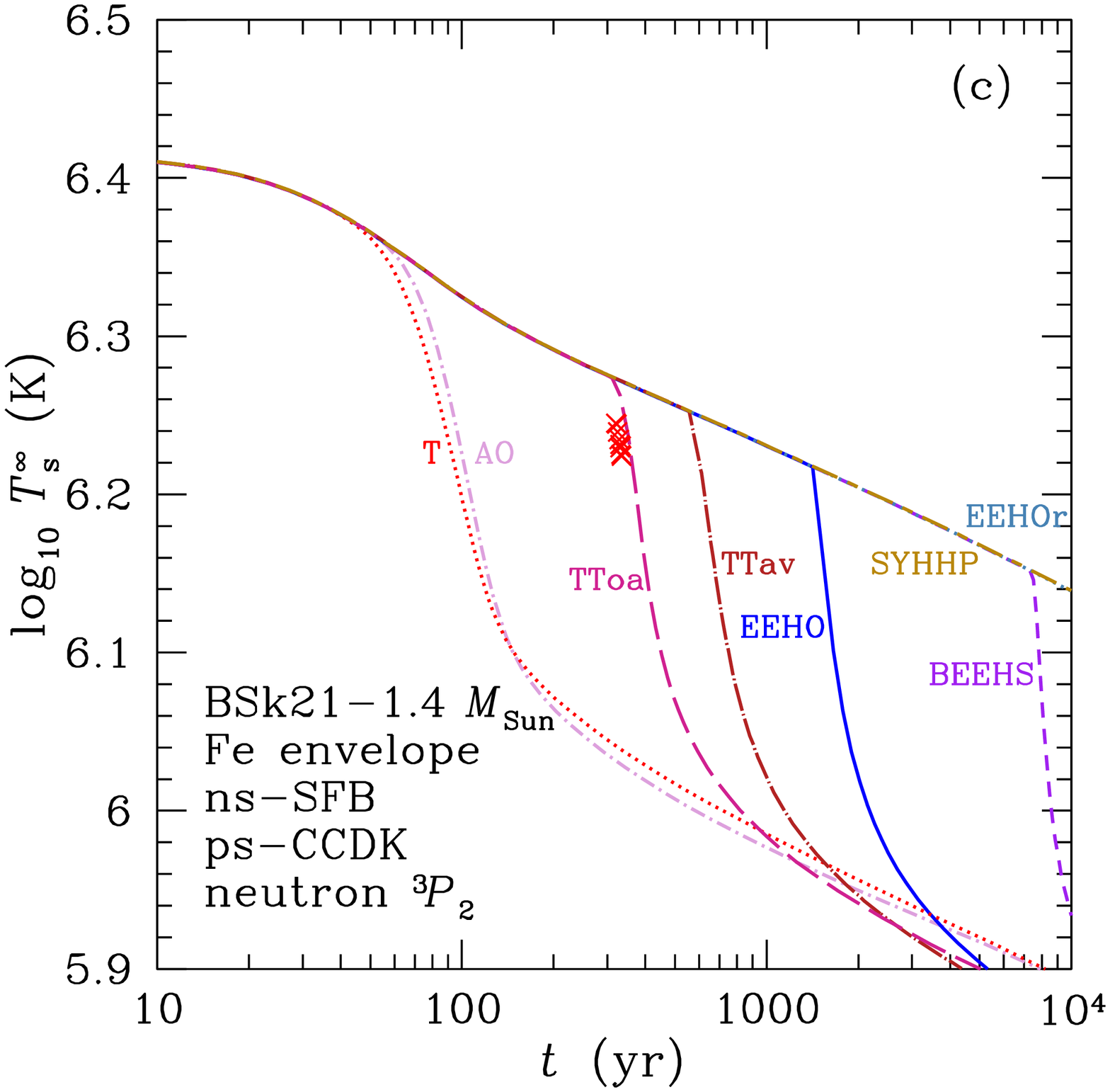}}
\caption{
(Color online)
Redshifted surface temperature $\Tsinfty$ as a function of age for a
$1.4\,\Msun$ NS using the APR (left), BSk20 (center), and BSk21 (right) EOS
with an iron envelope.
Different curves are cooling simulations using the SFB neutron singlet,
CCDK proton singlet, and one of various neutron triplet gap models
(see Fig.~\ref{fig:sfk}).
Crosses are the observed temperatures of the Cas~A NS.
\label{fig:cool}}
\end{figure*}

\begin{figure*}
\resizebox{\hsize}{!}{\includegraphics{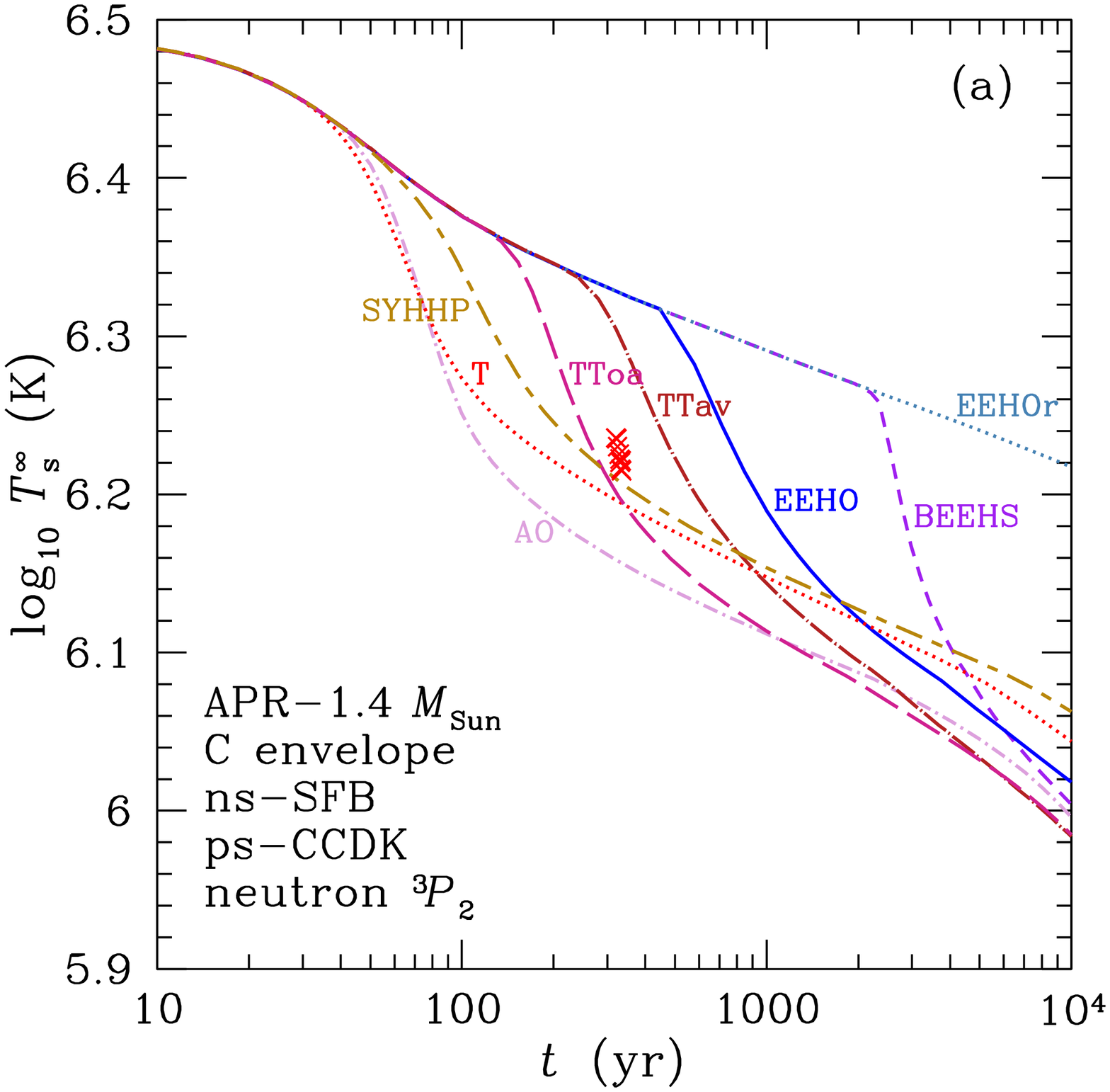} \includegraphics{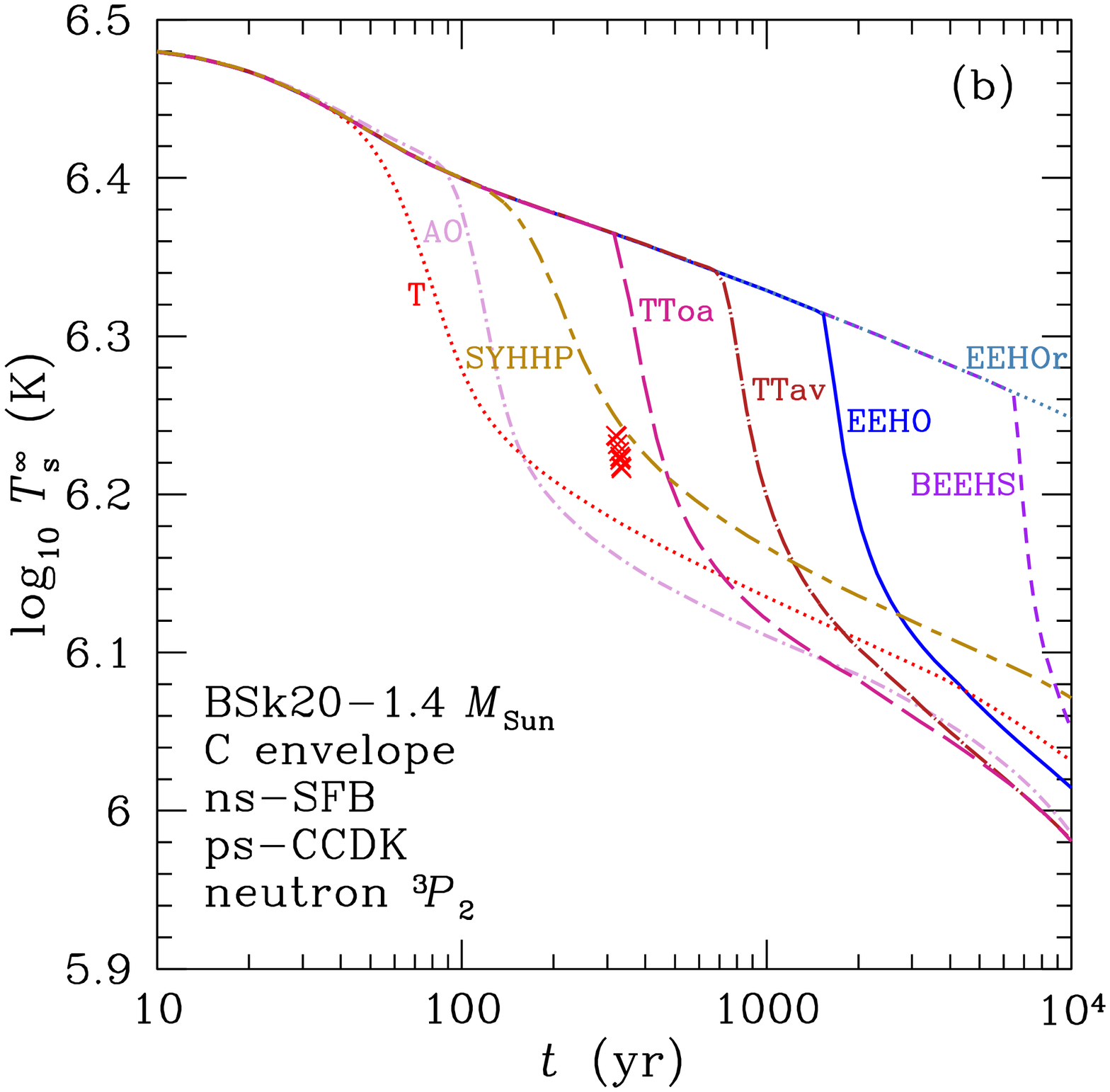} \includegraphics{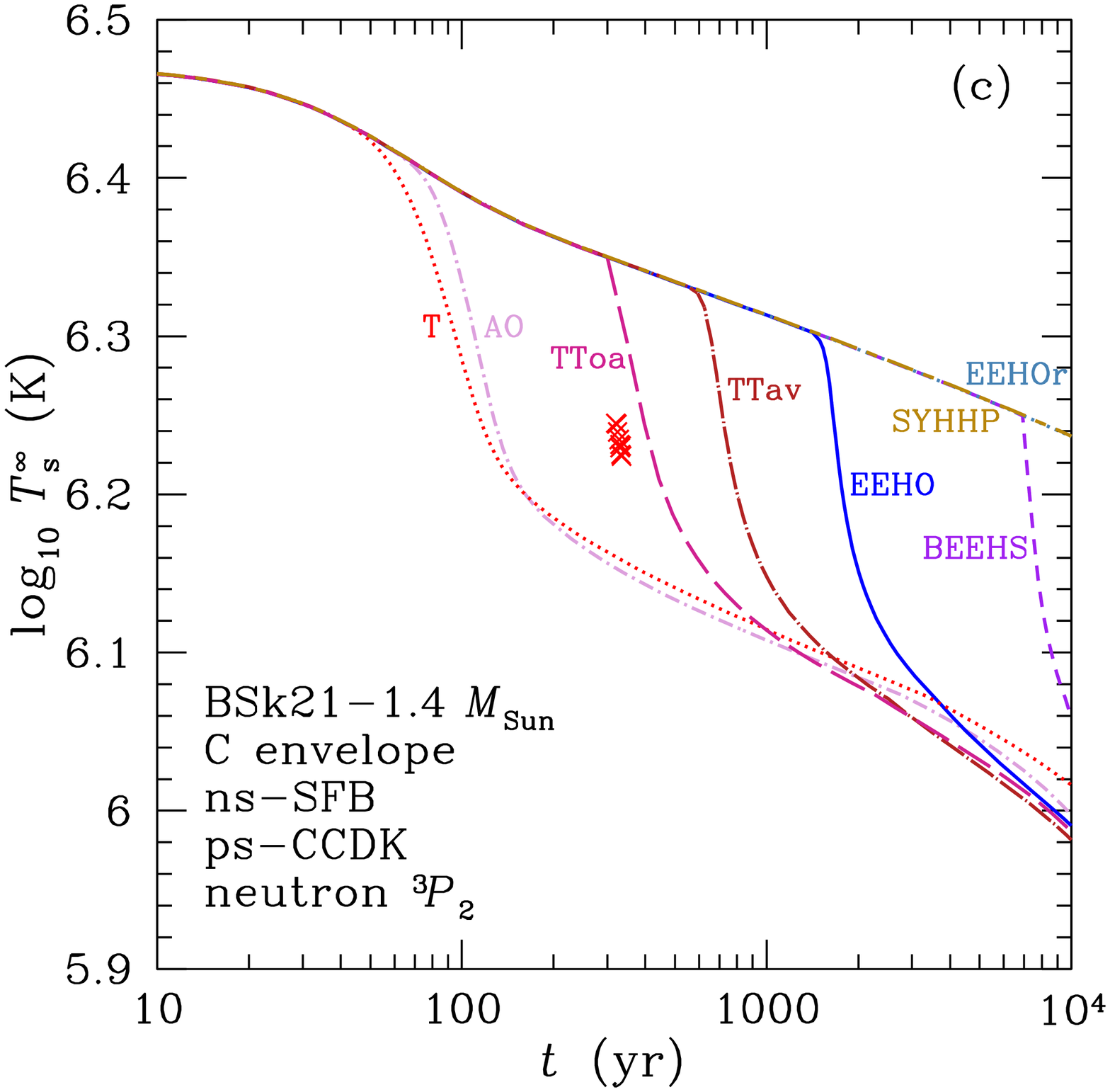}}
\caption{
(Color online)
Redshifted surface temperature $\Tsinfty$ as a function of age for a
$1.4\,\Msun$ NS using the APR (left), BSk20 (center), and BSk21 (right) EOS
with a carbon envelope ($\Delta M\approx 10^{-8}\Msun$).
Different curves are cooling simulations using the SFB neutron singlet,
CCDK proton singlet, and one of various neutron triplet gap models
(see Fig.~\ref{fig:sfk}).
Crosses are the observed temperatures of the Cas~A NS.
\label{fig:cool_acc8c}}
\end{figure*}

Figure~\ref{fig:cool} shows cooling curves calculated using a $1.4\,\Msun$
NS with the APR (left), BSk20 (center), and BSk21 (right) EOS and iron
envelope and including one neutron triplet gap model.
In all cases, we use the SFB model for the neutron singlet gap energy
and the CCDK model for the proton singlet gap energy.
Note that, for the NS ages of concern here, the EEHOr cooling curve is
identical to one from a NS model that has no neutron triplet superfluid.
Figure~\ref{fig:cool_acc8c} shows cooling curves calculated using the same
models as those used for the cooling curves of Fig.~\ref{fig:cool},
except for a maximally carbon-rich envelope ($\Delta M\approx 10^{-8}\Msun$).
We see that strong neutron triplet gaps produce temperature evolutions that
undergo an epoch of very rapid cooling (due to neutrino emission by Cooper
pair formation and breaking) once the temperature drops below the critical
temperature for the onset of superfluidity.
The time when this rapid cooling begins is strongly correlated with the
maximum of the critical temperature, i.e., earlier onset for a higher
temperature.  However, the density dependence of the critical temperature
is also important in determining initiation of rapid cooling
(see, e.g., model T versus AO versus SYHHP).
The variation of the critical temperature with density also determines
the rate of temperature decline since the fraction of the NS
that is becoming superfluid determines the neutrino luminosity.

%-------------------------------------------
%-------------------------------------------

%-------------------------------------------
%-------------------------------------------

\subsection{Fitting the Cas~A NS temperature evolution \label{sec:casa}}

We now test whether particular combinations of EOS and neutron triplet
gap models can fit the observed temperature decline of the Cas~A NS.
We again only consider the SFB model for neutron singlet and CCDK model
for proton singlet.  The former does not affect our results
(see Sec.~\ref{sec:ns}), while the latter is needed to sufficiently
suppress modified Urca processes prior to the current epoch of rapid
cooling (see Sec.~\ref{sec:ps}).
We consider either an iron envelope or a carbon envelope with
$\Delta M\sim 10^{-15}$, $10^{-11}$, or $10^{-8}\Msun$.

For each EOS and triplet gap model, we vary the NS mass (bearing in mind
the constraints obtained in Sec.~\ref{sec:fitsp} from fitting the Cas~A NS
spectra), calculate the temperature evolution, and note if the cooling curve
matches the observed temperature decline.
Once we find a potential match, we re-fit the spectra using the specific
NS mass and radius implied by the EOS under consideration, and then we
perform a least squares fit to the observed temperature decline.
Thus our derived mass and radius consistently fit both the spectra and
temperature evolution of the Cas~A NS.

\begin{figure}
\resizebox{0.95\hsize}{!}{\includegraphics{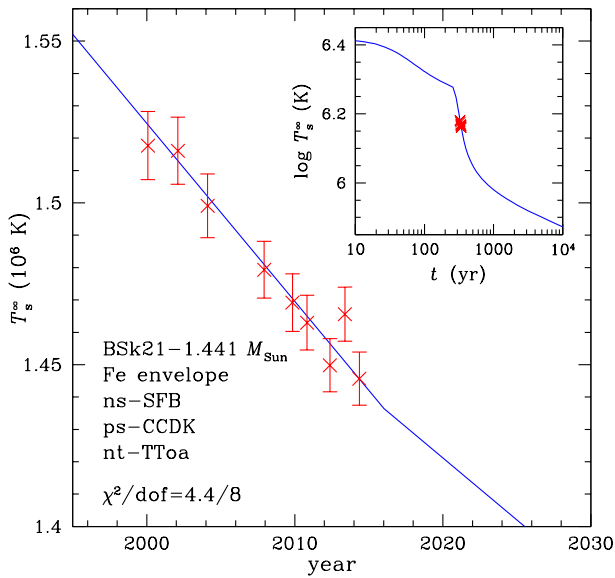}}
\caption{
(Color online)
Redshifted surface temperature $\Tsinfty$ as a function of year, with
redshift $1+\zg=1.229$.
Crosses and $1\sigma$ error bars are the observed \Chandra ACIS-S Graded
temperatures of the Cas~A NS.
Cooling curve is for a $M=1.441\,\Msun$ and $R=12.59\mbox{ km}$ NS built
using the BSk21 EOS with an iron envelope and SFB neutron singlet, CCDK
proton singlet, and TToa neutron triplet gap models.
Inset: Expanded view of temperature evolution as a function of time.
\label{fig:fitcool}}
\end{figure}

Despite the many possible combinations, we find only a few combinations that
match the observed spectra and cooling rate.
One solution yields
$M=1.812\,\Msun$ (BSk20 EOS, TToa triplet gap, iron envelope).
Other solutions yield
$M=1.582\,\Msun$ (BSk21 EOS, TTav triplet gap, iron envelope),
$M=1.441\,\Msun$ (BSk21 EOS, TToa triplet gap, iron envelope),
$M=1.441\,\Msun$ (BSk21 EOS, TToa triplet gap, carbon envelope with
$10^{-15}\Msun$), and
$M=1.582\,\Msun$ (BSk21 EOS, TToa triplet gap, carbon envelope with
$10^{-8}\Msun$).
Only three of these solutions give a good $\chisqr$ value for the least
squares fit of all the temperatures:
$\chisqr=0.55$ for $M=1.441\,\Msun$ and BSk21 EOS with iron envelope,
$\chisqr=0.47$ for $M=1.441\,\Msun$ and BSk21 EOS with $10^{-15}\Msun$
carbon envelope,
and $\chisqr=0.94$ for $M=1.812\,\Msun$ and BSk20 EOS with iron envelope,
all using the TToa triplet gap;
the fit also requires the supernova that produced the NS to have occurred
in the year 1674, 1669, and 1653, respectively, which matches well with the
determination from the expansion of the supernova remnant of $1681\pm19$
\cite{fesenetal06}.
The other two fits require the supernova to have occurred in the year
1617 and 1586, respectively.
We show the best-fit solution ($M=1.441\,\Msun$) in Fig.~\ref{fig:fitcool}.
Given the current systematic uncertainties, including absolute flux
calibration of the observations (see \cite{elshamoutyetal13,posseltetal13}),
we estimate a mass uncertainty of approximately $\sim 0.03\,\Msun$ for a
given EOS and gap model.

%-------------------------------------------
%-------------------------------------------

\section{Discussion \label{sec:discuss}}

For the first time, we successfully obtain consistent fits between the
nine epochs of \Chandra ACIS-S Graded spectra and the derived temperature
evolution.  Our best-fit yields a NS mass $M=1.44\,\Msun$ and radius
$R=12.6\mbox{ km}$ using the BSk21 EOS, TToa neutron triplet superfluid
and CCDK proton singlet superconductor gap models, and an iron envelope
or thin carbon layer (with $\Delta M\approx 10^{-15}\Msun$) on top of an
iron envelope.
Because there still exist large observational and theoretical uncertainties,
we cannot absolutely rule out the other EOSs or some of the other
superfluid and superconducting gap models considered here.
What we show is that it is possible to accurately measure the mass of a NS
using the method described.
Future work will examine what constraints are implied for the case where
the Cas~A NS is not cooling significantly or is cooling at a lower rate,
as suggested by the analyses of \cite{posseltetal13} and
\cite{elshamoutyetal13}, respectively.

While the parameterization of the gap energy [see Eq.~(\ref{eq:sfparam})]
is an approximation, we demonstrate the features that gap models should
possess if they are to fit the Cas~A NS observations.
In particular, the proton singlet gap should be large enough to permit
a large fraction of the core to become superconducting early in the age
of the NS in order to suppress early neutrino cooling.
The neutron triplet gap also needs to extend to a large fraction of the
core but with a maximum critical temperature that is just at the right
level so that rapid cooling does not initiate too early or too late in
order to explain the Cas~A observations \cite{pageetal11,shterninetal11}.
For the neutron singlet gap, its effect on the temperature evolution occurs
early on (age $\lesssim 10^2\mbox{ yr}$), during the thermal relaxation
phase when the NS interior is strongly non-isothermal \cite{pageetal09}.
The Cas~A data do not provide useful constraints for this gap.
However we note that some neutron singlet gap models (e.g., AWP2, SCLBL,
and SFB) extend beyond the inner crust into the core.  While such behavior
has no distinctive effect on the cooling behavior of an isolated NS, it
may affect observable phenomena such as pulsar glitches
\cite{anderssonetal12,chamel13,piekarewiczetal14,steineretal15}.

There are other possible explanations for the cooling behavior of the
Cas~A NS besides the onset of core superfluidity and superconductivity,
e.g., heating by r-mode oscillations \cite{yangetal11}
or magnetic field decay \cite{bonannoetal14},
very slow thermal relaxation \cite{blaschkeetal12},
rotationally-induced neutrino cooling \cite{negreirosetal13},
and transition to quark phases \cite{nodaetal13,sedrakian13}.
It would be interesting to see what constraints on some of these models
could be obtained by performing consistent fitting of the Cas~A spectra
and temperature evolution similar to the one performed here.

Finally, we note that it is desirable to use a single nuclear theory
calculation to obtain consistent EOS and superfluid and superconductor gap
energies.  However, this is not possible at the present time.
Our work is, in part, to motivate such a calculation.
A second purpose is to motivate the production of analytic approximations to
the detailed calculations performed by the
nuclear physics community, not just of the EOS [e.g., pressure as a function
of density $P(n)$], but also nucleon effective masses and superfluid and
superconducting gap energies [i.e., $m^{\mathrm{eff}}(n)$ and $\Delta(\kf)$].
Analytic approximations are vital for modeling of astrophysical sources,
and we note the valuable contributions of
\cite{haenselpotekhin04,chamel08} for SLy
and \cite{chameletal09,potekhinetal13} for BSk.

\begin{acknowledgments}
The authors are grateful to Dan Patnaude for providing the \Chandra data
and the anonymous referee for comments that led to improvements in the
manuscript.
W.C.G.H. is grateful to Dmitry Yakovlev for providing his cooling code and
invaluable assistance and advice.
W.C.G.H. thanks Dany Page for the APR EOS and Tatsuyuki Takatsuka for the
TTav and TToa gap models.
W.C.G.H. thanks Marcello Baldo, Alfio Bonanno, Fiorella Burgio,
and Hans-Josef Schulze for discussions.
W.C.G.H. appreciates use of computer facilities at the Kavli Institute for
Particle Astrophysics and Cosmology
and at Physics and Astronomy at University of Southampton.
W.C.G.H. acknowledges travel support to visit Universita di Catania from
NewCompStar, COST Action MP1304.
C.O.H. is supported by an Ingenuity New Faculty Award and a NSERC Discovery
Grant.
A.Y.P. is partly supported by the RFBR (grant~14-02-00868) and by the State
Program ``Leading Scientific Schools of RF'' (grant NSh~294.2014.2).
\end{acknowledgments}

%-------------------------------------------

\end{document}